\documentclass[article]{revtex4}
\begin{document}
\title{Detector Efficiency Limits on Quantum Improvement}
\author{Deborah J. Jackson}
\author{George M. Hockney}
\affiliation{%
Quantum Computing Technologies\\
 Jet Propulsion Laboratory, California Institute of Technology \\
4800 Oak Grove Drive, Pasadena, California 91109-8099 \\
}%
\date{\today}
\begin{abstract}
\begin{center}
\bf Although the National Institute of Standards and Technology has measured the intrinsic quantum efficiency of Si and InGaAs APD materials to be above $98\%$ by building an efficient compound detector, commercially available devices have efficiencies ranging between $15\%$ and $75\%$.  This means bandwidth, dark current, cost, and other factors are more important than quantum efficiency for existing applications.  For non-classical correlated photon applications, the system $SNR_{correlated} \propto \frac {\sqrt{\eta N}}{\sqrt(1 - \eta)}$,  rather than $SNR_{classical} \propto \sqrt{\eta N}$, which means the detector design trade space must be re-evaluated.  This paper systematically examines  the generic detection process, lays out the considerations needed for designing detectors for non-classical applications, and identifies the ultimate physical limits on quantum efficiency.   
\end{center}

~~~~~

\end{abstract}
\maketitle
\parskip 3pt

~~~~~

\section{Introduction}
Over the last decade many proposals have been made to improve S/N ratios in various devices using squeezed or entangled light\cite{LIGO}\cite{Kim}.  In all of these proposals, the main limiting factor is detector quantum efficiency.  Unlike systems using coherent light, where a change in detector efficiency from 0.5 to 0.9 would bring less than 3 dB S/N improvement, in entangled and squeezed systems the same change improves the S/N 
ratio by an order of magnitude.  For very similar reasons, linear optical quantum computing (LOQC)\cite{KLM} is
also directly limited by quantum efficiency.  This section explains in detail why this is  so, and why $1-\eta$, where $\eta$ is the detector efficiency, is a quantity of extreme importance.  Other difficulties, such as the inability to generate states with small Fano factors, are generally much less limiting than the detection quantum efficiency.

There are two general ways to use non-classical light in a device. The first, which has been used in most of the fundamental demonstrations (such as Mandel's demonstration of entangled interference\cite{Mandel}, or, more 
recently, teleportation\cite{tele}), is to use coincidence counting to extract only those events where no 
photon loss has occurred.  One very clever way to use this is the "heralded photon" method of measuring absolute detector efficiency\cite{Heralded}. In these systems detector inefficiency matters but the missed events are discarded by the coincidence veto.  While coincidence counting yields a clean sample, it does so by throwing away power, so that the overall power in the system is not reduced. The second, though, is the more interesting for achieving quantum improvements.  In the second sort of system, the intent is to reduce the overall power required for a given S/N ratio by using entangled or squeezed sources.  In these systems, $1 - \eta$ determines the possible improvement.  Its success is entirely dependent on one's ability to achieve high quantum efficiencies in the detection step.

To make a concrete example, consider Figure ~1(a)
\begin{figure}
\caption{(a) Upper spectrum is an example of the time dependent behavior of poissonian source and (b) shows the time dependent behavior of a perfectly squeezed source.}
\end{figure}
 which compares the behavior of a coherent source in the time domain with Poisson noise to a perfectly  correlated source with equally spaced photons.  Here we use time-domain correlation because it is straight-forward to analyze; however, the results apply to any correlation.  
The upper spectrum in  Figure 1(a) exhibits the time dependent behavior of a Poissonian source with shot noise fluctuations.  In contrast, the lower spectrum in Figure 1(a) shows the time-dependent behavior of a squeezed source whose  shot noise is almost eliminated.   Consider a signal $S$ which has $N$ photons from the coherent source.  This signal will have several sources of noise:

\begin{equation}
    \sigma^2 = {\sigma^2}_{shot} + {\sigma^2}_{dark counts} + 
{\sigma^2}_{stray} + {\sigma^2}_{etc}.
\end{equation}

However, in a well designed experiment, all the other sources of  noise fluctuations are reduced below the shot noise limit.  Assuming that we are in the regime where  ${\sigma^2}_{shot}$ is the limiting factor,

\begin{equation}
    \sigma^2 = {\sigma^2}_{shot} \propto N
\end{equation}

Therefore, $SNR = \frac{S}{\sqrt{\sigma^2}} \propto 
\frac{N}{\sqrt{N}} = \sqrt{N}$.  That is, the signal-to-noise ratio is proportional to the square root of the number of photons or the energy in the single shot, or the fluctuation in the signal is proportional to the square root of the power flowing through the system.

In contrast, with perfect correlation ${\sigma^2}_{shot} = 1$,  where the 1 comes from quantum uncertainty about the beginning and ending times.  In this case, we have

\begin{equation}
    \sigma^2 = \sigma^2_{shot} \propto 1
\end{equation}

and $SNR = \frac{S}{\sqrt{1}} \propto N$.  This means the  signal-to-noise is proportional to the power flowing through the system, and the fluctuation in the signal does not depend on the power.

This is the promise of entangled and squeezed light:  it can reduce the power required for a given $\frac{S}{N}$ by orders of magnitude if $N$ is the number of photons/sec in a typical laser beam.   However, this cannot be achieved in practice, and detector inefficiency is the most important limiting factor.

To see why this is so, consider Figure 1(b) which illustrates what happens if photons are lost from detection in an un-correlated way (which is intrinsic in all losses due to surface imperfections, absorption, and
detector inefficiency).  Then the remaining photons will have Poisson statistics at a level given by $1 - \eta$.  To make this precise, in a classical system the effect of a loss $\eta$ would be

\begin{equation}
    SNR \propto \frac{\eta N}{\sqrt{\eta N}} = \sqrt{\eta N},
\end{equation}

which is not significant if $\eta$ changes from, say, 0.9 to 0.99, but for the entangled case it introduces  fluctuations

\begin{equation}
    SNR \propto \frac{\eta N}{\sqrt{\eta +
                        (1-\eta)N}} \approx \frac {\sqrt{\eta 
N}}{\sqrt(1 - \eta)},
                      {\rm for} N \gg \frac{\eta}{1 - \eta}.
\end{equation}
For large $N$, this is only a factor of $\frac{1}{\sqrt(1-\eta)}$ better than the classical uncorrelated case. The full formula agrees with with the quantum Monte-Carlo calculations of Kim et. al. \cite{Kim}.  Since current detectors have values of $\eta$ below 0.9, this limits quantum improvements to a factor of about $\sqrt(10)$ for a given power level in the device.

The above expression is for perfectly squeezed light.  For real light, introduce the Fano factor$f$ such that ${\sigma^2}_{shot} = fN$, which characterizes the amount of squeezing.  The Fano factor varies from 1, for classical light, all the way to $1/N$ for perfectly squeezed light.  The expression for shot-noise limited systems with detector inefficiency $\eta$ is then given by

\begin{equation}
    \frac{S}{N} \propto \frac{\eta N}{\sqrt{\eta fN +
                        (1-\eta)N}} \approx \frac {\sqrt{\eta 
N}}{\sqrt{f + (1 - \eta)/\eta}},
\end{equation}
That is, the obtainable SNR is limited by imperfect squeezing and detector inefficiency, which both can make the denominator in equation [6] significantly larger than the ideal $\sqrt{1/N} $ of perfect squeezing.  In the limit where $f>>(1-\eta)/\eta$, the signal to noise ratio improves from the classical  $\sqrt{\eta N}$ proportional to $\sqrt{\eta N/f}$; and in the limit where  $f<<(1-\eta)/\eta$  the signal to noise ratio is given by the expression in equation [5].  Factors of up to 6dB over the shot noise limit have been achieved previously, which means 
$\sqrt{f + (1-\eta)/\eta}$ had to be less than 0.25.

\section{The Generic Detection Process}
As the discussion in the previous section has made clear, the ability to observe the unique effects due to the behavior of squeezed light is possible only with high quantum efficiency detection system.  This section examines the physical limits to getting the best achievable quantum efficiency using single photon 
sensitive detectors.   We refer to this as a "detection system", because proper observation of squeezed effects requires measurements that account for nearly all of the energy emitted by the source. Indeed Table I shows that the measured intrinsic quantum efficiency of high quality APD detector materials can be as high as 0.98.  Yet, the state of the art number quoted for off-the-shelf Si APD's hovers around 0.70.  To understand why, one must carefully examine all loss mechanisms associated with the detection process.  This implies that besides  a high intrinsic material quantum efficiency, a very efficient method of collecting the photons and capturing  them at the detector\ is needed.  The generic detection process consists of four steps:

	(1) photon collection

	(2) photon transmission into the absorber

	(3) photon conversion to a photo-electron

	(4) photo-electron multiplication

\begin{table}[h]
\caption{\label{material} Intrinsic Properties of Photodiode Materials \cite{Bachor}}
\begin{center}
\begin{tabular}{p{2.5 in}lp{1.2 in}lp{1.2 in}lp{1.2 in}}\hline
  -- \ & Si \ & Ge \  & InGaAs\\ \hline
Band \ gap \ $E_{BG}$ \ & 1.11 \ eV \ & 0.66 \ eV \ & ~ \ 1.0 \ eV  \\ \hline
  Peak \ detection \ wavelength, \ $\lambda_{peak}$ \ & 800 \ nm \ & 1600 \ nm \ &1000 \ nm \\ \hline
  Material \ quantum \ efficiency \ & 0.99 \ & 0.88 \ & 0.98 \\ \hline
Refractive \ index \ at \ $\lambda_{peak}$ \ & 3.5 \ & 4.0 \ & 3.7 \\  \hline
Normal \ incidence \ reflectivity \ & $31\%$ \ & $36\%$ \ & $33\%$  \\ \hline
Brewster's \ angle \ & $74^{o}$ \ & $75^{o}$ \ & $76^{o}$  \\ \hline

\end{tabular}
\end{center}
\end{table}
	
\emph{Photon collection efficiency - $\eta_{col}$}  Until recently, most single photon sensitive detection schemes depended on the ability to, first, create a photo electron, and then efficiently multiply that photoelectron to a large enough current pulse to register in an external circuit.  (Although the generic detection model described here assumes that the conversion step is always to a 
photo electron, there are a few approaches  SSPD, TES, and photon multipliers, where the amplification step uses another process to detect the presence of the photon.)  Each of these steps represents a decision point in the flow of the photon energy in that there are multiple paths or channels that open up for  the photon to take. For example, during the collection step, the photon is either coupled into the absorbing region of the detector or misses the absorber entirely.  Therefore any energy that is lost from reflections off of the collection optics or that is diffracted away from the detector  reduces the collection efficiency 
by a factor, $\eta_{col}$.  By choosing the dimensions of the source, detector, and collection optics to be large compared to the source wavelength, diffraction effects can be minimized.   The lowest losses are probably achievable with front surface optics and dielectric mirror films.   Also, the strategic use of anti-reflection coatings on the collection optics makes other losses at this stage negligible.  

\emph{Photon transmission into sample and absorption efficiency - $\eta_{abs}$}  Upon incidence at the 
absorbing surface of the detector, there are three paths for the final disposition of the photon: it is either  is reflected, transmitted, or absorbed in the detector.  Any photon reflected or completely transmitted through the detector represents energy lost from the detection process, further reducing the absorption efficiency by a factor, $ \eta_{abs}$.  It is possible to recover the reflected photons by adopting the light trapping geometry first introduced by Zalewski et. al. \cite{Zalewski} similar to the one in Figure~2.  
\begin{figure}
\caption{Example of a light trap detection scheme which utilizes three mirrors to get five bounces off of detector substrates. The table in the insert compares the quantum efficiency measured with a single bounce with that achieved by collecting the signal from a total of five bounces. }
\end{figure}

The trap concept assumes that the absorption depth is deep enough to  absorb $100\%$ of all photons entering the detector substrate (note: this will affect the readout bandwidth).  It works as follows.  Photons are introduced to a sequence of detectors where they are either absorbed or reflected. If the electronic outputs of all the detectors are summed and there are N detectors in the sequence, the geometry is such that an incoming photon will be reintroduced to the absorbing surface $2N-1$ times.  The net effect is to increase the absorption probability from $\eta_{abs}$ to 

\begin{equation}
   \eta_{eff-abs}=\sum_{n=1}^{2N-1}\eta_{abs}(1-\eta_{abs})^{n-1},         \label{eq:trap}
\end{equation}

\emph{Photon conversion to a photo-electron- $\eta_{p-e}$}  The conversion step may also open up other channels for disposing of the photon energy besides conversion to a photo-electron.  The presence of other energy channels at this step reduces the detector efficiency by a factor, $\eta_{p-e}$.  For example, the photon could be converted to vibrational energy that heats up the absorber 
instead, or there could be loss of minority carriers due to recombination.  Zalewski's data confirms that the principle loss mechanism in APD materials is recombination\cite{Zalewski}.  He classifies the loss mechanisms according to the three regions of the absorption material in which they occur,  (i) at the absorption interface, (ii) inside the absorption volume, (iii) outside the depletion region.  If the reverse bias on the photodiode extends the depletion region to the backside electrode, the third loss mechanism can be ignored.   For very pure and specially selected Si, Ge, and InGaAs detector bulk materials, high $ \eta_{p-e}$'s (up to 98$\%$) have been reported  \cite{Zalewski}, \cite{Bachor}.  These numbers were achieved using  the Zalewski's light trap geometry.    

\emph{Photo-electron multiplication- $\eta_{mul} \simeq 1$} The last factor, $\eta_{mul}$, reflects any mechanisms that contribute to missed counts during the  multiplication step.  For example, the detector dead time results in missed counts if a photon arrives before the detector has recovered from counting an earlier photon arrival.  This is the same effect as detector saturation.  Therefore, as long as the count rate $N_{c} << 1/\tau$, where  $\tau$ is the device dead time, $\eta_{mul} \simeq 1$, hence has negligible effect.  Thus the total quantum efficiency is a product of these four different processes, 

\begin{equation}
   \eta_{total} = \eta_{col} \eta_{eff-abs} \eta_{p-e} \eta_{mul}. 		\label{eq:QE}
\end{equation}

\section{Dark Counts and Saturation}

Besides shot noise,  the inherent noise sources in a  detector are thermal noise, dark current noise, and avalanche multiplier noise.   Table II compares the properties of detectors that are considered candidates for single photon sensitive detection.  All but the PMT and the APD are cooled to liquid helium temperatures so that the thermal noise contribution is negligible.   There are two basic types of single photon sensitive detectors.  The first type are those that operate in Geiger mode and can only detect one photon at a time.  This would include the  PMT, the APD,  and the Superconducting Single Photon Detector (SSPD) \cite{Verevkin}.  The second type are those that can count multiple photons simultaneously.  In this category are the the Visible Light Photon Counter (VLPC)\cite{Waks} and superconducting Tranistion Edge Sensor (TES)\cite{Miller} \cite{Karasik}.

The measured quantum efficiency can be obscured by other random processes which trigger the detector when the light from the source under test is blocked. These dark counts are  due to either stray light or noise currents intrinsic to the detector or else are caused by the readout circuitry.  Because one cannot count that portion of the signal that is obscured by the presence of dark counts, the measured quantum efficiency must be modified by adding  the dark count contribution

\begin{equation}
   \eta_{dark-adj} = (N_{c} - N_{d})/N = \eta_{total} - N_{d}/N. 
	\label{dark}
\end{equation}
where $N_{c}$ is the count rate per second of the detector, $N_{d}$ is the dark count rate, and $N$ is the photon flux, $N = (\lambda P)/hc$, where $P$ is the incident power and $hc/\lambda$ is the energy per photon.  Since stray light dark counts can be avoided through the careful design of the measurement testbed, we see that the intrinsic dark current of the detector itself  represents a hard limit on the the best quantum efficiency achievable by any given detector.  Ideally, the detector dark count should be less than the least significant figure in the quantum efficiency specification.  This means that if the applications requires 99.9$\%$ quantum efficiency, the dark count rate, $N_{d} \ < 0.001N$.  Therefore, the way to beat the dark count limit is to run the detector at the highest count rate possible. Another approach that works well with the VLPC is to reduce the effective dark count by gating the detector on for short time windows immediately after the source is triggered.  For the detectors listed in Table II, the dark currents for optimized TES and SSPD devices may not be intrinsically limited by the detector, but by the readout circuit noise.

 Devices that operate in Geiger mode must recover fully from processing one photon before they are ready to detect the next one.  Therefore the photon counting rate should not exceed $1/\tau_{D}$, where $\tau_{D}$ is the detector dead time. To keep the detector from saturating, one requires that $N_{C} <1/\tau_{D}$.  These two material properties of the detector therefore bracket the best counting rate and thus help determine a limit for the achievable quantum efficiencies. 

\begin{equation}
  1/\tau_{D} >> N_{c} >> N_{d} 
\end{equation}

In contrast, the saturation limit of photon counting detectors is higher.  It is determined by the size of the localized detecting area $a$ compared to the overall size of the detector aperture $A_{D}$.  When $a<<A_{D}$ the detector can be modeled as array with roughly $N_{E}\simeq A_{D}/a$ elements and the Kok-Braunstein model  \cite{Kok} can be used to calculate  the confidence that two simultaneously arriving photons do not cause a saturation event by striking the same spot.  This confidence level for a maximally entangled state is given by:

\begin{equation}
  C=\frac{N_{E}}{N_{E}+\delta[\eta^{2}+2N_{E}(1-\eta^{2})]} .   		\label{confidence}
\end{equation}

where $\delta=\alpha^{2}/2$, and $\alpha$ is defined as the mean expectation values of the number operator. 

\section{Photon Counting and Excess Noise Factor}

Besides the intrinsic noise due to the detector dark current, the noise from the amplification process also will affect the measurable quantum efficiency.  Note that the excess noise factor only impacts the quantum efficiency when one is operating in multiple photon counting mode.  Because it is a statistical process, any amplification introduces additional noise onto a signal.  McIntyre has developed theoretical models of the amplification process, and has reduced the quantification of the added noise to an excess noise factor, F, which is mathematically defined as 

\begin{equation}
F = \frac{\langle M^{2} \rangle}{\langle M 
\rangle^{2}} 
\end{equation}
where M is the number of electrons produced by a photo-ionization event.  This excess noise is a useful figure of merit for comparing different detector technologies.  It should be noted that F is greater than 1.0 for all of these detectors, but the VLPC, TES, and SSPD exhibit significant departures from the predictions of McIntyre's model.  Although it has not been measured for the TES and the SSPD detectors, the noise will clearly depend on the amplification electronics outside the detector.  Therefore, these two detectors, in particular, should have quantum limited excess noise factors when a high gain dc-SQUID array is used as the first stage amplifier.    





\begin{table}[h]
\caption{\label{SOA}Comparison of State-of-the-Art (SOA) Detector 
Performance Parameters}
\begin{center}
\begin{tabular}{p{1.1in}lp{1.0in}lp{1.0in}lp{2.0in}lp{1.0in}}\hline
   Technology\ & SOA \  BW \ & Dark \  Count \  & Operating \ Temp. \ & SOA \ QE \ &  ENF \\ \hline
PMT \ & 1.5 \ GHz \ & -- \ & 300K \ & 0.40 \ & 1.2 \\ \hline
  APD \ & 1 \ GHz \ & 25\ Hz \ &300K \ or \ 77 \ K \ & 0.75 \  & 2.0 \\ \hline
  VLPC \ & 300 \ MHz \ & 20 \ kHz \ &6K \ & 0.94 \ & 1.015 \\ \hline
  TES \ & 20 \ kHz \ & $\>$\ 0.001\ Hz \ & 0.1\ K \ &  0.20 \ & $\simeq$ 1 \\ \hline
SSPD \ & 30 \ GHz \ & $\>$\ 0.01\ Hz \ & 5 \ K \ & 0.03  \ & $\simeq$ \ 1 \\ 
    \ &     \ &     \ &     \ & (0.9 \ Light \ trap) \ &      \\  \hline
\end{tabular}
\end{center}
\end{table}
*Assumes $\eta_{det} = 0.125$. \\
Numbers in parentheses are either assumed values or calculated performance estimates.\\

\section{Conclusions}
Given the results of Zalewski et. al. \cite{Zalewski} showing that the intrinsic quantum efficiencies of Si and InGaAs APD materials are $99\%$ and $98\%$, respectively, one is led to question why the best commercial APD's report  quantum efficiencies between $15\%$ and $75\%$.   Clearly, $\eta_{p-e}$ is not the limiting factor!  In most cases, the lower $\eta_{total}$ is due to a combination of reflective losses (which are not controlled by light trapping) and transmission loss through the backside of the APD (whose thin absorption region insures higher bandwidth by reducing the transit time for sweeping carriers out of the absorption region).  Thus the architectural design requirements for  achieving the best quantum efficiency are being traded off against the design requirements for other performance parameters.    This means that improvements to the architecture of the detection system  which are aimed at altering the trade space are the key to realizing significant improvements in the quantum efficiency. 

An example of this can be illustrated by changing the Zalewski apparatus in Figure 2 to improve the bandwidth.  Because of the retro-reflecting end detector in the figure, it is not possible to precisely reconstruct the time at which the photon entered the device. If, however, one avoids retroreflecting the photons back down the detector chain by going to the geometry shown in Figure 3, 
\begin{figure}
\caption{This is an example of how the geometry in the figure above can be modified so that electronic timing can be used to increase the bandwidth of the detection system.  The $D_{n}$  signify delay lines that adjust the arrival of the detector signals so that they are summed simultaneously.}
\end{figure}
timing electronics can be used to reconstruct a higher bandwidth signal, at the expense of adding extra detectors.  

As this example shows, the whole \emph{detection system}  must be systematically optimized for minimum loss in order for high quantum efficiencies to be routinely measurable in quantum optics and quantum computing systems.  For example, because Zalewski et. al.'s \cite{Zalewski} intrinsic quantum efficiency measurements are suspiciously close to the state-of-the-art reflectivity of commercially available dielectric mirrors, they may have been limited by $\eta_{col}$, so more attention must be paid to insuring that the collection optics are not limiting.   

Achieving a detection quantum efficiency of $99.999\%$ is a very difficult objective which requires systematic elimination of any effects that mask the desired result.   We summarize by listing the improvements that must be realized in order to achieve an order of magnitude quantum efficiency improvement above the $99\%$ state-of-the-art measurement.\
\begin{list}
{\setlength{\rightmargin}{\leftmargin}}
\item (1) $\eta_{adj}$ --Ensure that the ratio of the source production rate to the dark count rate $N_{c}/N_{d}  >10^{3}$, thereby limiting the observation of saturation effects due to detector dead time.  From the data collected in Table II, it is clear that in order to use the VLPC for these types of measurements, the effective dark current, $N_{d-eff}$ must be reduced by gating ON for short time windows after the source is triggered.  
\item (2) $\eta_{mul}$ -- Eliminate saturation and dark count effects by using a count rate, $N_{c}$, that has the following upper and lower limits  $1/\tau_{D} >> N_{c} >> N_{d-eff}$. 
\item (3) $\eta_{col}$ --Ensure that the total losses from collection optics is less than $10^{-3}$. This is a very difficult objective.  The solution may be to find a low loss way to couple from the source to a light trap detector using all total internal reflection optics. Adiabatically tapered transitions between waveguides are one approach [10].
\item (4) $\eta_{eff-abs}$ --Adopt a light trap geometry.
\item (5) $\eta_{p-e}$ --ÐFor quantum efficiency measurements that adhere to the above four conditions, one should be able to directly observe the intrinsic quantum efficiency limit of the detection process in any given detection system. Therefore, the goal is to find the material with the highest intrinsic quantum efficiency. 
\end{list}

We expect that a careful and systematic treatment of the above suggestions will push the benchmark quantum efficiency above $99\%$.  Although we expect to see some improvement in the APD and VLPC systems, the detection mechanisms behind the TES and the SSPD devices are more likely to result in quantum efficiencies that start approaching $99.999\%$ because there are fewer leakage channels that can open up as loss channels during the detection event.

\acknowledgments
The research described in this paper was carried out at the Jet Propulsion Laboratory, California Institute of Technology.

\end{document}